\documentclass[letter]{aa}  

\usepackage{graphicx}
\usepackage{txfonts}
\usepackage[colorlinks=true,linkcolor=blue,citecolor=blue,urlcolor=blue]{hyperref}
\usepackage{physics}
\usepackage{amsmath}
\usepackage{xfrac}
\usepackage{float}
\usepackage{placeins}

\begin{document}

   \title{The fate of Earth during the Sun's giant phases}
   \subtitle{New constraints from ab initio tidal modelling and AGB mass loss}

   \author{ M. Esseldeurs\inst{1}\corrauth{mats.esseldeurs@kuleuven.be}\and 
            S. Mathis\inst{2}\email{stephane.mathis@cea.fr}\and
            L. Decin\inst{1}\email{leen.decin@kuleuven.be}}

   \institute{Instituut voor Sterrenkunde, KU Leuven, Celestijnenlaan 200D, 3001 Leuven, Belgium
         \and
             Université Paris-Saclay, Université Paris Cité, CEA, CNRS, AIM, 91191 Gif-sur-Yvette, France
             }

   \date{Received; accepted}

  \abstract
        {The long-term evolution of planetary systems around solar-type stars is governed by the interplay between stellar expansion, tidal interactions, and mass loss during the red giant branch (RGB) and asymptotic giant branch (AGB) phases. However, tidal dissipation efficiencies and AGB mass-loss rates both remain poorly constrained, leading to significant uncertainty in predicting the fate of planetary systems, in particular, that of the Earth orbiting the ageing Sun.}
        {We reassess the survival of the Earth and the inner Solar System planets during the post-main-sequence evolution of the Sun, focusing on the impact of updated tidal dissipation prescriptions and varying AGB mass-loss rates.}
        {We modelled the orbital evolution of the Earth using stellar evolution tracks for a solar-mass star that incorporate the most recent tidal dissipation prescriptions derived from ab initio calculations. We compared these results with outcomes obtained using previously published and commonly adopted tidal prescriptions, and we explored a range of AGB mass-loss rates.}
        {We find that the predicted fate of the Earth is highly sensitive to the tidal model and the assumed mass-loss rate. Based on updated tidal dissipation prescriptions, Earth survives the RGB and AGB phases of the Sun. In contrast, the use of earlier prescriptions of tidal dissipation prescriptions leads to engulfment during the AGB phase. Furthermore, low AGB mass-loss rates result in engulfment, and vice versa. Using the observed mass-loss rates of the AGB star L$_2$ Pup as a proxy for the Sun's future AGB mass-loss rate results in the survival of the Earth during the AGB phase when combined with our tidal dissipation evaluation.}
        {Currently, the survival of the Earth and the inner Solar System is not robustly determined and critically depends on the treatment of tidal dissipation and stellar mass loss. Given the current observational uncertainties in AGB mass-loss rates, the ultimate fate of the Earth remains uncertain, highlighting the need for improved constraints on the late-stages of stellar evolution. However, considering observational proxies for the Sun during the AGB phase, it is likely that the Earth will survive the Sun's red giant phase.}


   \maketitle

\section{Introduction}
    When stars deplete hydrogen in their core, they evolve off the main sequence (MS) and expand during the red giant branch (RGB) and asymptotic giant branch (AGB) phases \citep[e.g.][]{Kippenhahn1990}. During this evolution, the star grows in size and loses mass through a stellar wind. For the Sun, this expansion will result in a star that is larger than the current orbit of Earth (1.2--1.5 M$_\odot$ depending on the AGB mass-loss rate). This will have a significant impact on the orbital evolution of the Solar System planets. The fate of the inner Solar System planets, and in particular Earth, has been a topic of debate in the literature, with different studies reaching different conclusions. Some studies suggested that Earth will be engulfed by the expanding Sun during the RGB \citep[e.g.][]{Rybicki2001, Schroder2008, Lanza2023}, while others suggested that Earth will survive the RGB phase \citep[e.g.][]{Sackmann1993, Rasio1996, Nordhaus2010, Guo2016}. These studies often made different assumptions about the mass-loss rates and tidal interactions, which led to different conclusions. Additionally, some studies considered the impact of planet-planet interactions on the orbital evolution of the inner Solar System \citep[e.g.][]{Mogavero2021}, although they did not take the structural evolution of the Sun into account. Observationally, there is evidence of Earth-like planets around white dwarfs \citep[e.g.][]{Zhang2024}, which favours the survival of planets like Earth during the RGB and AGB phases. However, the exact fate of the Earth and the inner Solar System is still uncertain and requires further investigation.

    The effect of mass loss on the orbital evolution of the Solar System planets was first studied by \cite{Sackmann1993}, who assumed conservation of angular momentum when the star loses mass (effectively widening the orbit). \cite{Rybicki2001} analysed effects in addition to the mass loss, such as tidal interactions, mass accretion by the wind, evaporation by stellar irradiation, and wind drag. The authors found that mass loss and tidal interactions dominate the orbital evolution. \cite{Schroder2008} and \cite{Nordhaus2010} combined the effects of mass loss and tidal interactions and arrived at different conclusions (engulfment for \citealp{Schroder2008} and survival for \citealp{Nordhaus2010}). The difference in the conclusions of these studies can be attributed to the different assumptions made about the mass-loss rates during the RGB, where \cite{Nordhaus2010} assumed a higher RGB mass-loss rate than \cite{Schroder2008}. This dependence was further analysed by \cite{Guo2016}, who performed a detailed analysis of the effect of different mass-loss rates during the RGB on the orbital evolution. For low mass-loss rates during the RGB, the star not only spends more time in the RGB phase, but also more crucially, the star grows more significantly. This extends the time and enables stronger tidal interactions to shrink the orbit of the Earth, leading to engulfment, and vice versa. Based on the observationally calibrated RGB mass-loss rates of \cite{McDonald2015}, \cite{Guo2016} concluded that Earth will survive the RGB phase. \cite{Lanza2023} added the effect of residual eccentricity, which is a stochastic process in which density fluctuations around the planet lead to eccentricity fluctuations of about $10^{-5}$ during the MS that are enhanced to the order of $10^{-2}$ during the RGB, making the survival of the Earth a stochastic process.

    These previous studies either neglected the effect of tidal interactions altogether \cite[e.g.][]{Sackmann1993} or used simplified prescriptions for the tidal interactions based on the equilibrium tide theory of \cite{Zahn1966a,Zahn1977} \cite[e.g.][]{Rybicki2001, Schroder2008, Guo2016, Lanza2023}. These prescriptions are based on the assumption that the tidal dissipation is dominated by the equilibrium tide and use parametrisations for stellar properties. However, recent studies have improved these prescriptions by using ab initio modelling of the tidal dissipation based on the latest understanding of the internal structure and dynamics of evolved stars and including dynamical tides associated with progressive internal gravity waves \citep[see][and references therein]{Esseldeurs2024}. Along the same lines, mass loss during the AGB phase is also still uncertain \citep{Decin2021} and can affect the orbital evolution of planetary or binary systems significantly.

    We aim to improve the modelling of the tidal interactions by using the prescriptions of \cite{Esseldeurs2024} for the tidal dissipation. We also use the tools developed in \cite{EsseldeursDecin2026} to model the orbital evolution, which allows us to take the tidal interactions and the changes through stellar winds into account. Using these tools, we model the orbital evolution of the inner Solar System planets during the Sun's RGB and AGB phases and determine their fate during these phases.

    In Sect. \ref{sec:model} we describe the tools we used to model the orbital evolution of the inner Solar System. In Sect. \ref{sec:stellar_evolution} we outline the stellar evolutionary models we used for the Sun. In Sect. \ref{sec:orbital_evolution} we present the results of the orbital evolution of the inner Solar System and discuss the effect of different tidal dissipation prescriptions and AGB mass-loss rates on the fate of the Earth.

\section{Modelling the orbital evolution}\label{sec:model}
    To model the orbital evolution of the inner Solar System, we used the tools developed in \cite{EsseldeursDecin2026}. To follow changes in planetary orbital parameters, we considered tidal interactions in the Sun as well as stellar winds,
    \begin{equation}
        \frac{1}{a}\frac{\dd a}{\dd t} = \left.\frac{1}{a}\frac{\dd a}{\dd t}\right|_{\rm wind} + \left.\frac{1}{a}\frac{\dd a}{\dd t}\right|_{\rm tides} \ .
    \end{equation}
    For all simulations, the eccentricity remained negligible, but was taken into account for consistency. The equations for their evolution are given in Appendix \ref{app:eccentricity}.
    These differential equations were integrated for each planet independently, and we did not take the effect of planet-planet interactions into account. Earth only crosses the 5:3 mean motion resonance with Venus during the RGB phase, but this is expected to have only a minor effect on the orbital evolution of Earth (see Appendix \ref{app:meanmotionresonances}).

    The changes through stellar winds are given by
    \begin{equation}
        \left.\left\langle\frac{1}{a}\frac{\dd a}{\dd t}\right\rangle\right|_{\rm wind} \hspace{-5pt}= - 2 \frac{\dot m_1}{m_1}\left(1-\frac{\beta}{q} -  \eta (1-\beta)\frac{1+q}{q} - \frac{1-\beta}{2} \frac{1}{1+q}\right),
    \end{equation}
    where $\langle\rangle$ denotes an orbit-averaged quantity, $q$ is the mass fraction, $\eta$ is the specific angular momentum of material lost in units of the orbital angular momentum of the system per reduced mass, and $\beta = \frac{1+e^2}{(1-e^2)^{3/2}} \beta_\mathrm{c}$ is the enhanced mass-accretion efficiency effect accounting for the change in the orbital separation for eccentric orbits. $\eta$ and $\beta_\mathrm{c}$ were calculated using the analytical prescriptions proposed by \cite{Saladino2019a}.

    The changes through tides are given by
    \begin{equation}
        \left.\frac{1}{a}\frac{\dd a}{\dd t}\right|_\mathrm{tides} \hspace{-4pt}= - \frac{5}{4\pi} \frac{m_2}{m_1} {\left(\frac{R_\star}{a}\right)}^{5} \sum_{m=0}^{\infty} \sum_{n=-\infty}^{\infty} n\Omega_o |A_{2, m, n}(e)|^2 \Im (k^{2, m}_n)\ ,
    \end{equation}
    where $A_{2, m, n}(e)$ are the tidal coefficients (see \citealp{Ogilvie2014} table 1 and \citealp{EsseldeursDecin2026} for details), and $\Im (k^{2, m}_n)$ are the imaginary parts of the tidal Love numbers, which were calculated using the prescriptions of \cite{Esseldeurs2024}. These equations assume a negligible stellar rotation, which is a good approximation for evolved stars that have not been spun up through tidal interactions. This is a valid approximation for our Solar System calculations since tidal interactions are too weak to significantly spin the Sun up \citep{Madappatt2016}. The simulation either ended at the end of the stellar evolution model (a cool WD with $L = 10^{-1}$ L$_\odot$) or when the planet was engulfed by the star (when the orbital separation was smaller than the Roche radius).

\section{Stellar evolutionary models}\label{sec:stellar_evolution}
    We used the stellar evolutionary code MESA \citep{Paxton2011, Paxton2013, Paxton2015, Paxton2018, Paxton2019, Jermyn2023} to compute the evolution of the Sun from the pre-main sequence to the white dwarf phase. We used the same input physics as in \cite{Esseldeurs2024}, which were based on \cite{Cinquegrana2022} to be consistent with the Monash code \citep{Karakas2007} during the evolved phases. The initial mass of the Sun was set to $1$ M$_\odot$, and we used a metallicity of $Z=0.0134$ \citep{Asplund2009}. For the mass-loss rate during the RGB phase, we used the Reimers mass-loss rate prescription \citep{Reimers1975} with $\eta_\mathrm{Reimer}=0.477$, as calibrated from observations by \cite{McDonald2015}, and we used the Blöcker mass-loss rate prescription \citep{Blocker1995} with $\eta_\text{Bl{\"o}cker}$ between 0.01 and 0.5 (0.1 in \citealp{Esseldeurs2024}) for the mass-loss rate during the AGB phase, where 0.05 was taken as a reference value. A Kippenhahn diagram  is shown in Appendix \ref{app:Kippenhahn}.

\section{Orbital evolution predictions}\label{sec:orbital_evolution}
    \begin{figure}
        \centering
        \includegraphics[width=0.98\linewidth]{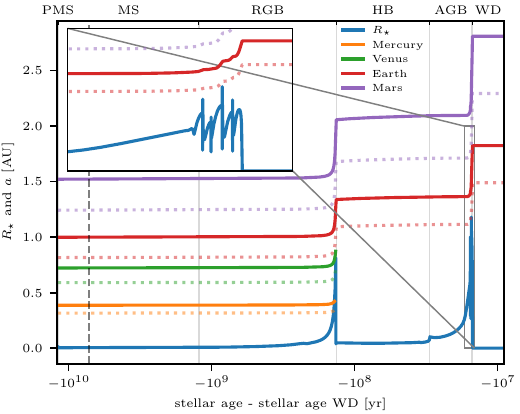}
        \vspace{-4pt}\caption{Orbital evolution (solid line, and the corresponding Solar Roche lobe radius, dotted line) of the inner Solar System during the evolved phases of the Sun. The different lines correspond to different planets (Mercury in orange, Venus in green, Earth in red, and Mars in purple). The current age of the Sun is indicated by the vertical dashed line.}\label{Fig:OrbitalEvolution}\vspace{-7pt}
    \end{figure}
    Using the stellar and orbital evolution models described in the previous section, we predicted the orbital evolution of the inner Solar System planets from the PMS until the end of the WD phase. The results are shown in Fig. \ref{Fig:OrbitalEvolution}, where we show the evolution of the orbital separation of the planets as a function of time until the WD phase. During the RGB phase, the orbital separation of the planets increases due to the mass loss of the star, where the increase is larger for planets that are farther away from the star. Mercury and Venus are not able to move outward fast enough to avoid engulfment and are engulfed during the RGB phase. Earth and Mars are able to move outward fast enough to avoid engulfment during the RGB phase. During the HB phase, the orbital separation of the planets stabilises as the star contracts, while during the AGB phase, the orbital separation of the planets increases again due to the mass loss of the star. During this phase, Earth and Mars are able to move outward fast enough to avoid engulfment. During the WD phase, the orbital separation of the planets stabilises again as the star contracts.

    \subsection{Impact of tidal dissipation modelling}
        \begin{figure}
            \centering
            \includegraphics[width=0.98\linewidth]{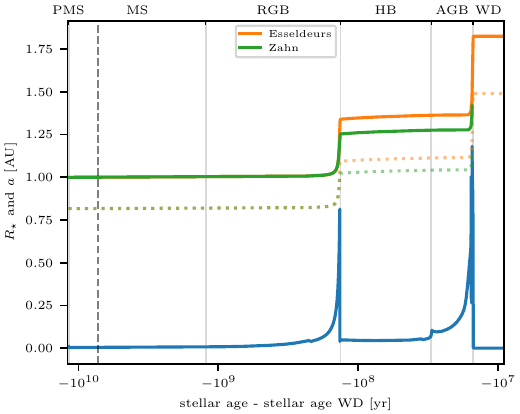}
            \vspace{-4pt}\caption{Orbital evolution of the Earth during the evolved phases of the Sun for different tidal dissipation prescriptions. In orange, we plot \cite{Esseldeurs2024} and in green, we plot \cite{Mustill2012} (based on \citealp{Zahn1966a}; see Appendix \ref{app:tidaldissipationequations}).}\vspace{-7pt}\label{Fig:OrbitalEvolution_TidalDissipation}
        \end{figure}
        To understand the effect of the different tidal dissipation prescriptions on the orbital evolution, we compared the results using the tidal prescriptions of \cite{Esseldeurs2024} to a \cite{Zahn1966a}-based prescription as used by \cite{Mustill2012} (see Appendix \ref{app:tidaldissipationequations} for a comparisson). The main difference between the prescriptions is the treatment of the frequency dependence of the turbulent viscosity. At Earth's orbit, the difference between the viscosity prescriptions is largest, making the choice of prescription crucial for Earth's orbital evolution.

        The results are shown in Fig. \ref{Fig:OrbitalEvolution_TidalDissipation}. The older prescriptions of \cite{Mustill2012} predict stronger tidal dissipation, which leads to a reduction in the orbital widening, and thus, to Earth leaving the RGB phase on a much closer orbit. Earth therefore starts at a shorter orbital distance at the start of the AGB phase, resulting in engulfment during the AGB phase. In contrast, our models predict Earth to be surviving the Sun's RGB and subsequent AGB evolution, assuming $\eta_\text{Bl{\"o}cker}=0.05$. As Earth is expected to move farther outward compared to previous studies, the stochastic eccentricity fluctuations induced by \cite{Lanza2023} are less important for Earth.

        Observationally, the tidal dissipation was tested in evolved stars using the circularisation of binary systems. In the RGB phase, \cite{Beck2018, Beck2022, Beck2024}, and \cite{Dewberry2025} have shown that the equilibrium tide dominates the tidal dissipation, but that the observed circularisation is stronger than predicted by the equilibrium tide using the \cite{Duguid2020} viscosity. In the AGB phase, \cite{EsseldeursDecin2026} have shown that the observed circularity of the binary system $\pi^1$ Gru is unexpected given current tidal predictions. This suggests that tidal dissipation in evolved stars might be stronger than predicted by current models, making Earth's engulfment more likely. However, these observations are based on binary circularisation, which involves higher-frequency tides than circular migration of orbits such as in the Earth-Sun system, and it might therefore not be directly applicable.\vspace{-11pt}

    \subsection{Impact of mass-loss rate prescriptions}
        In the past, the importance of mass loss during the RGB and AGB phases was highlighted by several studies \citep[e.g.][]{Sackmann1993,Guo2016}. \cite{Guo2016} investigated the effect of different $\eta_\text{Reimers}$ values (for describing the mass-loss rate during the RGB phase) on the orbital evolution and found that for values below $\eta_\text{Reimers} = 0.46$, Earth is engulfed during the RGB phase, and vice versa. As our tidal dissipation is weaker, this value will decrease when recomputed using our model. Because $\eta_\text{Reimers}$ is relatively well constrained by observations as $\eta_\mathrm{Reimer}=0.477\pm0.07$ \citep{McDonald2015}, the fate of the Earth during the RGB phase is relatively well constrained, and the Earth is expected to survive the RGB phase, but the engulfment is still within the uncertainty.

        The AGB mass-loss rates are more uncertain. Different prescriptions predict mass-loss rates that differ by more than one order of magnitude \citep{Decin2021}. Additionally, within the same prescription, different values of the free parameter $\eta_\text{Bl{\"o}cker}$ are used in the literature. To constrain this parameter observationally for our Sun, we considered the observations of the AGB star L$_2$ Pup, which is an AGB star with an initial mass close to the Sun's (0.98 M$_\odot$; \citealp{Kervella2016}) and can thus be used as a proxy for the Sun during the AGB phase. L$_2$ Pup is surrounded by a dusty disk, inside which a potential planet of mass $12\pm16$ M$_\text{Jup}$ was detected \citep{Kervella2016}. There have been two different estimates of the mass-loss rate of L$_2$ Pup, one estimate based on the dust emission, which gives a mass-loss rate of $\dot m_1 = 1 \times 10^{-6}$ M$_\odot$ yr$^{-1}$ \citep{Haworth2018}, and the other estimate based on the CO emission, which gives a much lower mass-loss rate of $\dot m_1 = 1.2 \times 10^{-8}$ M$_\odot$ yr$^{-1}$ \citep{Hoai2022}. We refer to Appendix \ref{app:mass-loss} for a discussion, but the large difference between the two estimates already highlights the uncertainty in the AGB mass-loss rates. With the Blöcker mass-loss rate prescription, the dust-based mass-loss rate corresponds to $\eta_\text{Bl{\"o}cker} = 2.5$, while the CO-based mass-loss rate corresponds to $\eta_\text{Bl{\"o}cker} = 0.03$. Values between 0.01 and 1 are commonly used in the literature. We took $\eta_\text{Bl{\"o}cker} = 0.05$ as a reference value (see Appendix \ref{app:mass-loss}), but we also explored the effect of different values between 0.01 and 0.5 for $\eta_\text{Bl{\"o}cker}$ on the orbital evolution of the Earth. It is worth noting that other AGB mass-loss prescriptions exist, such as the period-based prescription of \citet{Vassiliadis1993}, which cannot be directly compared to the luminosity-based Blöcker prescription. However, $\eta_\text{Bl{\"o}cker}$ can be calibrated to reproduce a similar mass-loss evolution within the explored parameter space.
        \begin{figure}
            \centering
            \includegraphics[width=0.98\linewidth]{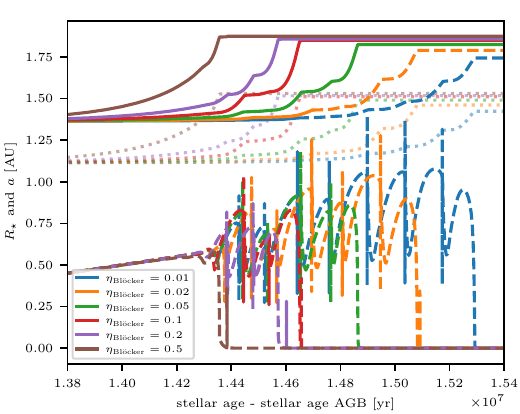}
            \vspace{-6pt}\caption{Orbital evolution of the Earth during the evolved phases of the Sun for different AGB mass-loss rates. The different lines correspond to different values for $\eta_\text{Bl{\"o}cker}$. The dashed lines correspond to the Solar radius, and the dotted lines correspond to the Solar Roche lobe radius. After the radius of the Sun exceeds the Roche lobe radius, the solid line for the orbital distance is changed to a dashed line to highlight the uncertainty in the model after this point (see text).}\vspace{-10pt}\label{Fig:OrbitalEvolution_MassLoss}
        \end{figure}

        The orbital evolution of Earth during the thermally pulsing AGB phase for different $\eta_\text{Bl{\"o}cker}$ values is shown in Fig. \ref{Fig:OrbitalEvolution_MassLoss}. For low $\eta_\text{Bl{\"o}cker}$ values (0.01 and 0.02), the Sun's radius exceeds the Roche lobe radius. Under normal circumstances, this would lead to Earth's engulfment. However, this occurs during a thermal pulse lasting only a few hundred years, making it unclear whether the interaction is strong enough to cause engulfment. For $\eta_\text{Bl{\"o}cker} = 0.01$, the Solar radius grows nearly as large as Earth's orbit, making engulfment likely. For $\eta_\text{Bl{\"o}cker} = 0.02$ (and 0.03, not shown), the Solar radius exceeds the Roche lobe radius only slightly (filling factor of 7\%), making the outcome uncertain. To anccurately model this would require stellar evolution calculations including the star-planet interaction during the thermally pulsing AGB phase (e.g. MESA in BINARY mode; \citealp{Paxton2015}), combined with an analytical model of the orbital evolution due to mass transfer \citep[e.g.][]{Parkosidis2026a, Parkosidis2026b}, which is beyond the scope of this paper. For $\eta_\text{Bl{\"o}cker} \geq 0.04$, the Solar radius remains below the Roche lobe radius, allowing Earth to survive the AGB phase. Because the AGB mass-loss rates are uncertain, Earth's fate remains uncertain, although the observational values of L$_2$ Pup suggest that Earth will likely survive the AGB phase.
    
\section{Conclusion}
    Our results show that the fate of Earth and the inner Solar System during the Sun's evolved phases strongly depends on tidal interactions and AGB mass-loss rates. Using updated tidal dissipation prescriptions based on ab initio modelling of evolved stars, we found that Earth survives the RGB and AGB phases. In contrast, previous tidal prescriptions predicted engulfment during the AGB phase. We also found that low AGB mass-loss rates lead to engulfment, while high rates allow Earth to survive. Since the AGB mass-loss rates remain observationally uncertain, the final fate of Earth is still unclear. However, the observed properties of L$_2$ Pup, which we used as a proxy for the future Sun, suggest that Earth will likely survive the AGB phase. Future observational constraints, combining spatially resolved interferometric data with the most recent hydrochemical simulations, are needed to better constrain AGB mass-loss rates and hence the fate of the inner Solar System. Additionally, the number of detected planets around red giants is expected to increase substantially in the coming years, particularly with the PLATO mission \citep{Rauer2025}. This will enable us to conduct population studies of the planetary orbital evolution around evolved stars and help us to constrain the future evolution of the Earth-Sun system.\vspace*{-4pt}

\begin{acknowledgements}
    The authors would  like to thank the anonymous referee for their constructive comments which helped to improve the quality of the paper.
    M. Esseldeurs, S. Mathis and L. Decin acknowledge support from the FWO grant G0B3823N. M. Esseldeurs and L. Decin acknowledge support from the FWO grant G099720N, the KU Leuven C1 excellence grant MAESTRO C16/17/007, the KU Leuven IDN grant ESCHER IDN/19/028 and the KU Leuven methusalem SOUL grant METH/24/012. S. Mathis acknowledges support from the PLATO CNES grant at CEA/DAp, from the Programme National de Planétologie (PNP-CNRS/INSU) and from the European Research Council through HORIZON ERC SyG Grant 4D-STAR 101071505. L. Decin acknowledges support from the FWO sabbatical grant K803625N. While partially funded by the European Union, views and opinions expressed are however those of the author only and do not necessarily reflect those of the European Union or the European Research Council. Neither the European Union nor the granting authority can be held responsible for them.\vspace*{-20pt}
\end{acknowledgements}

\bibliographystyle{aa}
\bibliography{bibliography}

@ARTICLE{EsseldeursDecin2026,
       author = {{Esseldeurs}, Mats and {Decin}, Leen and {De Ridder}, Joris and {Mori}, Yoshiya and {Karakas}, Amanda I. and {Malfait}, Jolien and {Danilovich}, Ta{\'\i}ssa and {Mathis}, St{\'e}phane and {Richards}, Anita M.~S. and {Sahai}, Raghvendra and {Yates}, Jeremy and {Van de Sande}, Marie and {Baes}, Maarten and {Baudry}, Alain and {Bolte}, Jan and {Ceulemans}, Thomas and {De Ceuster}, Frederik and {El Mellah}, Ileyk and {Etoka}, Sandra and {Gottlieb}, Carl and {Herpin}, Fabrice and {Kervella}, Pierre and {Landri}, Camille and {Marinho}, Louise and {McDonald}, Iain and {Menten}, Karl and {Millar}, Tom and {Osborn}, Zara and {Pimpanuwat}, Bannawit and {Plane}, John and {Price}, Daniel J. and {Siess}, Lionel and {Vermeulen}, Owen and {Wong}, Ka Tat},
        title = "{Evidence for the Keplerian orbit of a close companion around a giant star}",
      journal = {Nature Astronomy},
         year = 2026,
        month = jan,
       volume = {10},
        pages = {124-143},
          doi = {10.1038/s41550-025-02697-2},
archivePrefix = {arXiv},
       eprint = {2511.11247},
 primaryClass = {astro-ph.SR},
       adsurl = {https://ui.adsabs.harvard.edu/abs/2026NatAs..10..124E}
}

@ARTICLE{Saladino2019a,
       author = {{Saladino}, M.~I. and {Pols}, O.~R. and {Abate}, C.},
        title = "{Slowly, slowly in the wind. 3D hydrodynamical simulations of wind mass transfer and angular-momentum loss in AGB binary systems}",
      journal = {\aap},
         year = 2019,
        month = jun,
       volume = {626},
          eid = {A68},
        pages = {A68},
          doi = {10.1051/0004-6361/201834598},
archivePrefix = {arXiv},
       eprint = {1903.04515},
 primaryClass = {astro-ph.SR},
       adsurl = {https://ui.adsabs.harvard.edu/abs/2019A&A...626A..68S}
}

@ARTICLE{Ogilvie2014,
       author = {{Ogilvie}, Gordon I.},
        title = "{Tidal Dissipation in Stars and Giant Planets}",
      journal = {\araa},
         year = 2014,
        month = aug,
       volume = {52},
        pages = {171-210},
          doi = {10.1146/annurev-astro-081913-035941},
archivePrefix = {arXiv},
       eprint = {1406.2207},
 primaryClass = {astro-ph.SR},
       adsurl = {https://ui.adsabs.harvard.edu/abs/2014ARA&A..52..171O}
}

@ARTICLE{Esseldeurs2024,
       author = {{Esseldeurs}, M. and {Mathis}, S. and {Decin}, L.},
        title = "{Tidal dissipation in evolved low- and intermediate-mass stars}",
      journal = {\aap},
         year = 2024,
        month = oct,
       volume = {690},
          eid = {A266},
        pages = {A266},
          doi = {10.1051/0004-6361/202449648},
archivePrefix = {arXiv},
       eprint = {2407.10573},
 primaryClass = {astro-ph.SR},
       adsurl = {https://ui.adsabs.harvard.edu/abs/2024A&A...690A.266E}
}

@ARTICLE{Madappatt2016,
       author = {{Madappatt}, Niyas and {De Marco}, Orsola and {Villaver}, Eva},
        title = "{The effect of tides on the population of PN from interacting binaries}",
      journal = {\mnras},
         year = 2016,
        month = nov,
       volume = {463},
       number = {1},
        pages = {1040-1056},
          doi = {10.1093/mnras/stw2025},
archivePrefix = {arXiv},
       eprint = {1608.03041},
 primaryClass = {astro-ph.SR},
       adsurl = {https://ui.adsabs.harvard.edu/abs/2016MNRAS.463.1040M}
}

@ARTICLE{Paxton2011,
       author = {{Paxton}, Bill and {Bildsten}, Lars and {Dotter}, Aaron and {Herwig}, Falk and {Lesaffre}, Pierre and {Timmes}, Frank},
        title = "{Modules for Experiments in Stellar Astrophysics (MESA)}",
      journal = {\apjs},
         year = 2011,
        month = jan,
       volume = {192},
       number = {1},
          eid = {3},
        pages = {3},
          doi = {10.1088/0067-0049/192/1/3},
archivePrefix = {arXiv},
       eprint = {1009.1622},
 primaryClass = {astro-ph.SR},
       adsurl = {https://ui.adsabs.harvard.edu/abs/2011ApJS..192....3P}
}

@ARTICLE{Paxton2013,
       author = {{Paxton}, Bill and {Cantiello}, Matteo and {Arras}, Phil and {Bildsten}, Lars and {Brown}, Edward F. and {Dotter}, Aaron and {Mankovich}, Christopher and {Montgomery}, M.~H. and {Stello}, Dennis and {Timmes}, F.~X. and {Townsend}, Richard},
        title = "{Modules for Experiments in Stellar Astrophysics (MESA): Planets, Oscillations, Rotation, and Massive Stars}",
      journal = {\apjs},
         year = 2013,
        month = sep,
       volume = {208},
       number = {1},
          eid = {4},
        pages = {4},
          doi = {10.1088/0067-0049/208/1/4},
archivePrefix = {arXiv},
       eprint = {1301.0319},
 primaryClass = {astro-ph.SR},
       adsurl = {https://ui.adsabs.harvard.edu/abs/2013ApJS..208....4P}
}

@ARTICLE{Paxton2015,
       author = {{Paxton}, Bill and {Marchant}, Pablo and {Schwab}, Josiah and {Bauer}, Evan B. and {Bildsten}, Lars and {Cantiello}, Matteo and {Dessart}, Luc and {Farmer}, R. and {Hu}, H. and {Langer}, N. and {Townsend}, R.~H.~D. and {Townsley}, Dean M. and {Timmes}, F.~X.},
        title = "{Modules for Experiments in Stellar Astrophysics (MESA): Binaries, Pulsations, and Explosions}",
      journal = {\apjs},
         year = 2015,
        month = sep,
       volume = {220},
       number = {1},
          eid = {15},
        pages = {15},
          doi = {10.1088/0067-0049/220/1/15},
archivePrefix = {arXiv},
       eprint = {1506.03146},
 primaryClass = {astro-ph.SR},
       adsurl = {https://ui.adsabs.harvard.edu/abs/2015ApJS..220...15P}
}

@ARTICLE{Paxton2018,
       author = {{Paxton}, Bill and {Schwab}, Josiah and {Bauer}, Evan B. and {Bildsten}, Lars and {Blinnikov}, Sergei and {Duffell}, Paul and {Farmer}, R. and {Goldberg}, Jared A. and {Marchant}, Pablo and {Sorokina}, Elena and {Thoul}, Anne and {Townsend}, Richard H.~D. and {Timmes}, F.~X.},
        title = "{Modules for Experiments in Stellar Astrophysics (MESA): Convective Boundaries, Element Diffusion, and Massive Star Explosions}",
      journal = {\apjs},
         year = 2018,
        month = feb,
       volume = {234},
       number = {2},
          eid = {34},
        pages = {34},
          doi = {10.3847/1538-4365/aaa5a8},
archivePrefix = {arXiv},
       eprint = {1710.08424},
 primaryClass = {astro-ph.SR},
       adsurl = {https://ui.adsabs.harvard.edu/abs/2018ApJS..234...34P}
}

@ARTICLE{Paxton2019,
       author = {{Paxton}, Bill and {Smolec}, R. and {Schwab}, Josiah and {Gautschy}, A. and {Bildsten}, Lars and {Cantiello}, Matteo and {Dotter}, Aaron and {Farmer}, R. and {Goldberg}, Jared A. and {Jermyn}, Adam S. and {Kanbur}, S.~M. and {Marchant}, Pablo and {Thoul}, Anne and {Townsend}, Richard H.~D. and {Wolf}, William M. and {Zhang}, Michael and {Timmes}, F.~X.},
        title = "{Modules for Experiments in Stellar Astrophysics (MESA): Pulsating Variable Stars, Rotation, Convective Boundaries, and Energy Conservation}",
      journal = {\apjs},
         year = 2019,
        month = jul,
       volume = {243},
       number = {1},
          eid = {10},
        pages = {10},
          doi = {10.3847/1538-4365/ab2241},
archivePrefix = {arXiv},
       eprint = {1903.01426},
 primaryClass = {astro-ph.SR},
       adsurl = {https://ui.adsabs.harvard.edu/abs/2019ApJS..243...10P}
}

@ARTICLE{Jermyn2023,
       author = {{Jermyn}, Adam S. and {Bauer}, Evan B. and {Schwab}, Josiah and {Farmer}, R. and {Ball}, Warrick H. and {Bellinger}, Earl P. and {Dotter}, Aaron and {Joyce}, Meridith and {Marchant}, Pablo and {Mombarg}, Joey S.~G. and {Wolf}, William M. and {Sunny Wong}, Tin Long and {Cinquegrana}, Giulia C. and {Farrell}, Eoin and {Smolec}, R. and {Thoul}, Anne and {Cantiello}, Matteo and {Herwig}, Falk and {Toloza}, Odette and {Bildsten}, Lars and {Townsend}, Richard H.~D. and {Timmes}, F.~X.},
        title = "{Modules for Experiments in Stellar Astrophysics (MESA): Time-dependent Convection, Energy Conservation, Automatic Differentiation, and Infrastructure}",
      journal = {\apjs},
         year = 2023,
        month = mar,
       volume = {265},
       number = {1},
          eid = {15},
        pages = {15},
          doi = {10.3847/1538-4365/acae8d},
archivePrefix = {arXiv},
       eprint = {2208.03651},
 primaryClass = {astro-ph.SR},
       adsurl = {https://ui.adsabs.harvard.edu/abs/2023ApJS..265...15J}
}

@ARTICLE{Cinquegrana2022,
       author = {{Cinquegrana}, Giulia C. and {Joyce}, Meridith},
        title = "{Solar Calibration of the Convective Mixing Length for Use with the {\AE}SOPUS Opacities in MESA}",
      journal = {Research Notes of the American Astronomical Society},
         year = 2022,
        month = apr,
       volume = {6},
       number = {4},
          eid = {77},
        pages = {77},
          doi = {10.3847/2515-5172/ac6611},
       adsurl = {https://ui.adsabs.harvard.edu/abs/2022RNAAS...6...77C}
}

@INCOLLECTION{Reimers1975,
       author = {{Reimers}, D.},
        title = "{Circumstellar envelopes and mass loss of red giant stars.}",
    booktitle = {Problems in stellar atmospheres and envelopes.},
         year = 1975,
        pages = {229-256},
     publisher = {Springer-Verlag New York},
       adsurl = {https://ui.adsabs.harvard.edu/abs/1975psae.book..229R}
}

@ARTICLE{McDonald2015,
       author = {{McDonald}, I. and {Zijlstra}, A.~A.},
        title = "{Mass-loss on the red giant branch: the value and metallicity dependence of Reimers' {\ensuremath{\eta}} in globular clusters}",
      journal = {\mnras},
         year = 2015,
        month = mar,
       volume = {448},
       number = {1},
        pages = {502-521},
          doi = {10.1093/mnras/stv007},
archivePrefix = {arXiv},
       eprint = {1501.00874},
 primaryClass = {astro-ph.SR},
       adsurl = {https://ui.adsabs.harvard.edu/abs/2015MNRAS.448..502M}
}

@ARTICLE{Blocker1995,
       author = {{Bl{\"o}cker}, T.},
        title = "{Stellar evolution of low and intermediate-mass stars. I. Mass loss on the AGB and its consequences for stellar evolution.}",
      journal = {\aap},
         year = 1995,
        month = may,
       volume = {297},
        pages = {727},
       adsurl = {https://ui.adsabs.harvard.edu/abs/1995A&A...297..727B}
}

@ARTICLE{Zahn1966a,
       author = {{Zahn}, J.~P.},
        title = "{Les mar{\'e}es dans une {\'e}toile double serr{\'e}e}",
      journal = {Annales d'Astrophysique},
         year = 1966,
        month = feb,
       volume = {29},
        pages = {313},
       adsurl = {https://ui.adsabs.harvard.edu/abs/1966AnAp...29..313Z}
}

@ARTICLE{Zahn1977,
       author = {{Zahn}, J. -P.},
        title = "{Tidal friction in close binary systems.}",
      journal = {\aap},
         year = 1977,
        month = may,
       volume = {57},
        pages = {383-394},
       adsurl = {https://ui.adsabs.harvard.edu/abs/1977A&A....57..383Z}
}

@ARTICLE{Goldreich1977,
       author = {{Goldreich}, P. and {Keeley}, D.~A.},
        title = "{Solar seismology. II. The stochastic excitation of the solar p-modes by turbulent convection.}",
      journal = {\apj},
         year = 1977,
        month = feb,
       volume = {212},
        pages = {243-251},
          doi = {10.1086/155043},
       adsurl = {https://ui.adsabs.harvard.edu/abs/1977ApJ...212..243G}
}

@ARTICLE{Hut1981,
       author = {{Hut}, P.},
        title = "{Tidal evolution in close binary systems.}",
      journal = {\aap},
         year = 1981,
        month = jun,
       volume = {99},
        pages = {126-140},
       adsurl = {https://ui.adsabs.harvard.edu/abs/1981A&A....99..126H}
}

@ARTICLE{Hurley2002,
       author = {{Hurley}, Jarrod R. and {Tout}, Christopher A. and {Pols}, Onno R.},
        title = "{Evolution of binary stars and the effect of tides on binary populations}",
      journal = {\mnras},
         year = 2002,
        month = feb,
       volume = {329},
       number = {4},
        pages = {897-928},
          doi = {10.1046/j.1365-8711.2002.05038.x},
archivePrefix = {arXiv},
       eprint = {astro-ph/0201220},
 primaryClass = {astro-ph},
       adsurl = {https://ui.adsabs.harvard.edu/abs/2002MNRAS.329..897H}
}

@ARTICLE{Duguid2020,
       author = {{Duguid}, Craig D. and {Barker}, Adrian J. and {Jones}, C.~A.},
        title = "{Convective turbulent viscosity acting on equilibrium tidal flows: new frequency scaling of the effective viscosity}",
      journal = {\mnras},
         year = 2020,
        month = sep,
       volume = {497},
       number = {3},
        pages = {3400-3417},
          doi = {10.1093/mnras/staa2216},
archivePrefix = {arXiv},
       eprint = {2007.12624},
 primaryClass = {astro-ph.EP},
       adsurl = {https://ui.adsabs.harvard.edu/abs/2020MNRAS.497.3400D}
}

@ARTICLE{Schroder2008,
       author = {{Schr{\"o}der}, K.-P. and {Smith}, Robert Connon},
        title = "{Distant future of the Sun and Earth revisited}",
      journal = {\mnras},
         year = 2008,
        month = may,
       volume = {386},
       number = {1},
        pages = {155-163},
          doi = {10.1111/j.1365-2966.2008.13022.x},
archivePrefix = {arXiv},
       eprint = {0801.4031},
 primaryClass = {astro-ph},
       adsurl = {https://ui.adsabs.harvard.edu/abs/2008MNRAS.386..155S}
}

@ARTICLE{Sackmann1993,
       author = {{Sackmann}, I.-Juliana and {Boothroyd}, Arnold I. and {Kraemer}, Kathleen E.},
        title = "{Our Sun. III. Present and Future}",
      journal = {\apj},
         year = 1993,
        month = nov,
       volume = {418},
        pages = {457},
          doi = {10.1086/173407},
       adsurl = {https://ui.adsabs.harvard.edu/abs/1993ApJ...418..457S}
}

@ARTICLE{Nordhaus2010,
       author = {{Nordhaus}, J. and {Spiegel}, D.~S. and {Ibgui}, L. and {Goodman}, J. and {Burrows}, A.},
        title = "{Tides and tidal engulfment in post-main-sequence binaries: period gaps for planets and brown dwarfs around white dwarfs}",
      journal = {\mnras},
         year = 2010,
        month = oct,
       volume = {408},
       number = {1},
        pages = {631-641},
          doi = {10.1111/j.1365-2966.2010.17155.x},
archivePrefix = {arXiv},
       eprint = {1002.2216},
 primaryClass = {astro-ph.SR},
       adsurl = {https://ui.adsabs.harvard.edu/abs/2010MNRAS.408..631N}
}

@ARTICLE{Rybicki2001,
       author = {{Rybicki}, K.~R. and {Denis}, C.},
        title = "{On the Final Destiny of the Earth and the Solar System}",
      journal = {\icarus},
         year = 2001,
        month = may,
       volume = {151},
       number = {1},
        pages = {130-137},
          doi = {10.1006/icar.2001.6591},
       adsurl = {https://ui.adsabs.harvard.edu/abs/2001Icar..151..130R}
}

@ARTICLE{Guo2016,
       author = {{Guo}, Jianpo and {Lin}, Ling and {Bai}, Chunyan and {Liu}, Jinzhong},
        title = "{The effects of solar Reimers {\ensuremath{\eta}} on the final destinies of Venus, the Earth, and Mars}",
      journal = {\apss},
         year = 2016,
        month = apr,
       volume = {361},
          eid = {122},
        pages = {122},
          doi = {10.1007/s10509-016-2684-5},
       adsurl = {https://ui.adsabs.harvard.edu/abs/2016Ap&SS.361..122G}
}

@ARTICLE{Rasio1996,
       author = {{Rasio}, F.~A. and {Tout}, C.~A. and {Lubow}, S.~H. and {Livio}, M.},
        title = "{Tidal Decay of Close Planetary Orbits}",
      journal = {\apj},
         year = 1996,
        month = oct,
       volume = {470},
        pages = {1187},
          doi = {10.1086/177941},
archivePrefix = {arXiv},
       eprint = {astro-ph/9605059},
 primaryClass = {astro-ph},
       adsurl = {https://ui.adsabs.harvard.edu/abs/1996ApJ...470.1187R}
}

@ARTICLE{Mogavero2021,
       author = {{Mogavero}, Federico and {Laskar}, Jacques},
        title = "{Long-term dynamics of the inner planets in the Solar System}",
      journal = {\aap},
         year = 2021,
        month = nov,
       volume = {655},
          eid = {A1},
        pages = {A1},
          doi = {10.1051/0004-6361/202141007},
archivePrefix = {arXiv},
       eprint = {2105.14976},
 primaryClass = {astro-ph.EP},
       adsurl = {https://ui.adsabs.harvard.edu/abs/2021A&A...655A...1M}
}

@ARTICLE{Lanza2023,
       author = {{Lanza}, A.~F. and {Lebreton}, Y. and {Sallard}, C.},
        title = "{Residual eccentricity of an Earth-like planet orbiting a red giant Sun}",
      journal = {\aap},
         year = 2023,
        month = jun,
       volume = {674},
          eid = {A176},
        pages = {A176},
          doi = {10.1051/0004-6361/202345860},
archivePrefix = {arXiv},
       eprint = {2304.07808},
 primaryClass = {astro-ph.EP},
       adsurl = {https://ui.adsabs.harvard.edu/abs/2023A&A...674A.176L}
}

@ARTICLE{Zhang2024,
       author = {{Zhang}, Keming and {Zang}, Weicheng and {El-Badry}, Kareem and {Lu}, Jessica R. and {Bloom}, Joshua S. and {Agol}, Eric and {Gaudi}, B. Scott and {Konopacky}, Quinn and {LeBaron}, Natalie and {Mao}, Shude and {Terry}, Sean},
        title = "{An Earth-mass planet and a brown dwarf in orbit around a white dwarf}",
      journal = {Nature Astronomy},
         year = 2024,
        month = dec,
       volume = {8},
        pages = {1575-1582},
          doi = {10.1038/s41550-024-02375-9},
archivePrefix = {arXiv},
       eprint = {2409.02157},
 primaryClass = {astro-ph.EP},
       adsurl = {https://ui.adsabs.harvard.edu/abs/2024NatAs...8.1575Z}
}

@ARTICLE{Mustill2012,
       author = {{Mustill}, Alexander J. and {Villaver}, Eva},
        title = "{Foretellings of Ragnar{\"o}k: World-engulfing Asymptotic Giants and the Inheritance of White Dwarfs}",
      journal = {\apj},
         year = 2012,
        month = dec,
       volume = {761},
       number = {2},
          eid = {121},
        pages = {121},
          doi = {10.1088/0004-637X/761/2/121},
archivePrefix = {arXiv},
       eprint = {1210.0328},
 primaryClass = {astro-ph.EP},
       adsurl = {https://ui.adsabs.harvard.edu/abs/2012ApJ...761..121M}
}

@ARTICLE{Decin2021,
       author = {{Decin}, Leen},
        title = "{Evolution and Mass Loss of Cool Ageing Stars: a Daedalean Story}",
      journal = {\araa},
         year = 2021,
        month = sep,
       volume = {59},
        pages = {337-389},
          doi = {10.1146/annurev-astro-090120-033712},
archivePrefix = {arXiv},
       eprint = {2011.13472},
 primaryClass = {astro-ph.SR},
       adsurl = {https://ui.adsabs.harvard.edu/abs/2021ARA&A..59..337D}
}

@ARTICLE{Asplund2009,
       author = {{Asplund}, Martin and {Grevesse}, Nicolas and {Sauval}, A. Jacques and {Scott}, Pat},
        title = "{The Chemical Composition of the Sun}",
      journal = {\araa},
         year = 2009,
        month = sep,
       volume = {47},
       number = {1},
        pages = {481-522},
          doi = {10.1146/annurev.astro.46.060407.145222},
archivePrefix = {arXiv},
       eprint = {0909.0948},
 primaryClass = {astro-ph.SR},
       adsurl = {https://ui.adsabs.harvard.edu/abs/2009ARA&A..47..481A}
}

@ARTICLE{Kervella2016,
       author = {{Kervella}, P. and {Homan}, W. and {Richards}, A.~M.~S. and {Decin}, L. and {McDonald}, I. and {Montarg{\`e}s}, M. and {Ohnaka}, K.},
        title = "{ALMA observations of the nearby AGB star L$_{2}$ Puppis. I. Mass of the central star and detection of a candidate planet}",
      journal = {\aap},
         year = 2016,
        month = dec,
       volume = {596},
          eid = {A92},
        pages = {A92},
          doi = {10.1051/0004-6361/201629877},
archivePrefix = {arXiv},
       eprint = {1611.06231},
 primaryClass = {astro-ph.SR},
       adsurl = {https://ui.adsabs.harvard.edu/abs/2016A&A...596A..92K}
}

@ARTICLE{Vassiliadis1993,
       author = {{Vassiliadis}, E. and {Wood}, P.~R.},
        title = "{Evolution of Low- and Intermediate-Mass Stars to the End of the Asymptotic Giant Branch with Mass Loss}",
      journal = {\apj},
         year = 1993,
        month = aug,
       volume = {413},
        pages = {641},
          doi = {10.1086/173033},
       adsurl = {https://ui.adsabs.harvard.edu/abs/1993ApJ...413..641V}
}

@ARTICLE{Kervella2014,
       author = {{Kervella}, P. and {Montarg{\`e}s}, M. and {Ridgway}, S.~T. and {Perrin}, G. and {Chesneau}, O. and {Lacour}, S. and {Chiavassa}, A. and {Haubois}, X. and {Gallenne}, A.},
        title = "{An edge-on translucent dust disk around the nearest AGB star, L$_{2}$ Puppis. VLT/NACO spectro-imaging from 1.04 to 4.05 {\ensuremath{\mu}}m and VLTI interferometry}",
      journal = {\aap},
         year = 2014,
        month = apr,
       volume = {564},
          eid = {A88},
        pages = {A88},
          doi = {10.1051/0004-6361/201323273},
archivePrefix = {arXiv},
       eprint = {1404.3189},
 primaryClass = {astro-ph.SR},
       adsurl = {https://ui.adsabs.harvard.edu/abs/2014A&A...564A..88K}
}

@ARTICLE{Bedding2002,
       author = {{Bedding}, T.~R. and {Zijlstra}, A.~A. and {Jones}, A. and {Marang}, F. and {Matsuura}, M. and {Retter}, A. and {Whitelock}, P.~A. and {Yamamura}, I.},
        title = "{The light curve of the semiregular variable L$_{2}$ Puppis - I. A recent dimming event from dust}",
      journal = {\mnras},
         year = 2002,
        month = nov,
       volume = {337},
       number = {1},
        pages = {79-86},
          doi = {10.1046/j.1365-8711.2002.05822.x},
archivePrefix = {arXiv},
       eprint = {astro-ph/0207485},
 primaryClass = {astro-ph},
       adsurl = {https://ui.adsabs.harvard.edu/abs/2002MNRAS.337...79B}
}

@ARTICLE{Olofsson2002,
       author = {{Olofsson}, H. and {Gonz{\'a}lez Delgado}, D. and {Kerschbaum}, F. and {Sch{\"o}ier}, F.~L.},
        title = "{Mass loss rates of a sample of irregular and semiregular M-type AGB-variables}",
      journal = {\aap},
         year = 2002,
        month = sep,
       volume = {391},
        pages = {1053-1067},
          doi = {10.1051/0004-6361:20020841},
archivePrefix = {arXiv},
       eprint = {astro-ph/0206172},
 primaryClass = {astro-ph},
       adsurl = {https://ui.adsabs.harvard.edu/abs/2002A&A...391.1053O}
}

@ARTICLE{Jura2002,
       author = {{Jura}, M. and {Chen}, C. and {Plavchan}, P.},
        title = "{The Very Slow Wind from the Pulsating Semiregular Red Giant, L$_{2}$ Puppis}",
      journal = {\apj},
         year = 2002,
        month = apr,
       volume = {569},
       number = {2},
        pages = {964-974},
          doi = {10.1086/339207},
archivePrefix = {arXiv},
       eprint = {astro-ph/0112230},
 primaryClass = {astro-ph},
       adsurl = {https://ui.adsabs.harvard.edu/abs/2002ApJ...569..964J}
}

@ARTICLE{Danilovich2015,
       author = {{Danilovich}, T. and {Teyssier}, D. and {Justtanont}, K. and {Olofsson}, H. and {Cerrigone}, L. and {Bujarrabal}, V. and {Alcolea}, J. and {Cernicharo}, J. and {Castro-Carrizo}, A. and {Garc{\'\i}a-Lario}, P. and {Marston}, A.},
        title = "{New observations and models of circumstellar CO line emission of AGB stars in the Herschel SUCCESS programme}",
      journal = {\aap},
         year = 2015,
        month = sep,
       volume = {581},
          eid = {A60},
        pages = {A60},
          doi = {10.1051/0004-6361/201526705},
archivePrefix = {arXiv},
       eprint = {1506.09065},
 primaryClass = {astro-ph.SR},
       adsurl = {https://ui.adsabs.harvard.edu/abs/2015A&A...581A..60D}
}

@ARTICLE{Haworth2018,
       author = {{Haworth}, Thomas J. and {Booth}, Richard A. and {Homan}, Ward and {Decin}, Leen and {Clarke}, Cathie J. and {Mohanty}, Subhanjoy},
        title = "{Radiation-pressure-driven sub-Keplerian rotation of the disc around the AGB star L$_{2}$ Pup}",
      journal = {\mnras},
         year = 2018,
        month = jan,
       volume = {473},
       number = {1},
        pages = {317-327},
          doi = {10.1093/mnras/stx2416},
archivePrefix = {arXiv},
       eprint = {1709.02815},
 primaryClass = {astro-ph.SR},
       adsurl = {https://ui.adsabs.harvard.edu/abs/2018MNRAS.473..317H}
}

@ARTICLE{Hoai2022,
       author = {{Hoai}, D.~T. and {Nhung}, P.~T. and {Darriulat}, P. and {Diep}, P.~N. and {Ngoc}, N.~B. and {Thai}, T.~T. and {Tuan-Anh}, P.},
        title = "{Morpho-kinematics of the wind of asymptotic giant branch star L$_{2}$ Pup}",
      journal = {\mnras},
         year = 2022,
        month = feb,
       volume = {510},
       number = {2},
        pages = {2363-2378},
          doi = {10.1093/mnras/stab3465},
archivePrefix = {arXiv},
       eprint = {2111.13303},
 primaryClass = {astro-ph.SR},
       adsurl = {https://ui.adsabs.harvard.edu/abs/2022MNRAS.510.2363H}
}

@ARTICLE{Uttenthaler2024,
       author = {{Uttenthaler}, S.},
        title = "{The evolutionary state of the red giant star L$_{2}$ Puppis}",
      journal = {\aap},
         year = 2024,
        month = dec,
       volume = {692},
          eid = {A224},
        pages = {A224},
          doi = {10.1051/0004-6361/202452173},
archivePrefix = {arXiv},
       eprint = {2411.13388},
 primaryClass = {astro-ph.SR},
       adsurl = {https://ui.adsabs.harvard.edu/abs/2024A&A...692A.224U}
}

@ARTICLE{Parkosidis2026a,
       author = {{Parkosidis}, A. and {Toonen}, S. and {Dosopoulou}, F. and {Laplace}, E.},
        title = "{Rethinking mass transfer: A unified semianalytical framework for circular and eccentric binaries: I. Orbital evolution due to conservative mass transfer}",
      journal = {\aap},
         year = 2026,
        month = feb,
       volume = {706},
          eid = {A79},
        pages = {A79},
          doi = {10.1051/0004-6361/202555096},
archivePrefix = {arXiv},
       eprint = {2509.05243},
 primaryClass = {astro-ph.SR},
       adsurl = {https://ui.adsabs.harvard.edu/abs/2026A&A...706A..79P}
}

@ARTICLE{Parkosidis2026b,
       author = {{Parkosidis}, A. and {Toonen}, S. and {Laplace}, E. and {Dosopoulou}, F.},
        title = "{Rethinking mass transfer: A unified semianalytical framework for circular and eccentric binaries: II. Orbital evolution due to nonconservative mass transfer}",
      journal = {\aap},
         year = 2026,
        month = feb,
       volume = {706},
          eid = {A357},
        pages = {A357},
          doi = {10.1051/0004-6361/202558055},
archivePrefix = {arXiv},
       eprint = {2511.07190},
 primaryClass = {astro-ph.SR},
       adsurl = {https://ui.adsabs.harvard.edu/abs/2026A&A...706A.357P}
}

@ARTICLE{Karakas2007,
       author = {{Karakas}, Amanda and {Lattanzio}, John C.},
        title = "{Stellar Models and Yields of Asymptotic Giant Branch Stars}",
      journal = {\pasa},
         year = 2007,
        month = oct,
       volume = {24},
       number = {3},
        pages = {103-117},
          doi = {10.1071/AS07021},
archivePrefix = {arXiv},
       eprint = {0708.4385},
 primaryClass = {astro-ph},
       adsurl = {https://ui.adsabs.harvard.edu/abs/2007PASA...24..103K}
}

@BOOK{Kippenhahn1990,
       author = {{Kippenhahn}, Rudolf and {Weigert}, Alfred},
        title = "{Stellar Structure and Evolution}",
         year = 1990,
    publisher = {Springer-Verlag Berlin Heidelberg New York},
       adsurl = {https://ui.adsabs.harvard.edu/abs/1990sse..book.....K}
}

@ARTICLE{Rauer2025,
       author = {{Rauer}, Heike and {Aerts}, Conny and {Cabrera}, Juan and {Deleuil}, Magali and {Erikson}, Anders and {Gizon}, Laurent and {Goupil}, Mariejo and {Heras}, Ana and {Walloschek}, Thomas and {Lorenzo-Alvarez}, Jose and {Marliani}, Filippo and {Martin-Garcia}, C{\'e}sar and {Mas-Hesse}, J. Miguel and {O'Rourke}, Laurence and {Osborn}, Hugh and {Pagano}, Isabella and {Piotto}, Giampaolo and {Pollacco}, Don and {Ragazzoni}, Roberto and {Ramsay}, Gavin and {Udry}, St{\'e}phane and {Appourchaux}, Thierry and {Benz}, Willy and {Brandeker}, Alexis and {G{\"u}del}, Manuel and {Janot-Pacheco}, Eduardo and {Kabath}, Petr and {Kjeldsen}, Hans and {Min}, Michiel and {Santos}, Nuno and {Smith}, Alan and {Suarez}, Juan-Carlos and {Werner}, Stephanie C. and {Aboudan}, Alessio and {Abreu}, Manuel and {Acu{\~n}a}, Lorena and {Adams}, Moritz and {Adibekyan}, Vardan and {Affer}, Laura and {Agneray}, Fran{\c{c}}ois and {Agnor}, Craig and {Aguirre B{\o}rsen-Koch}, Victor and {Ahmed}, Saad and {Aigrain}, Suzanne and {Al-Bahlawan}, Ashraf and {Alcacera Gil}, Ma de los Angeles and {Alei}, Eleonora and {Alencar}, Silvia and {Alexander}, Richard and {Alfonso-Garz{\'o}n}, Julia and {Alibert}, Yann and {Allende Prieto}, Carlos and {Almeida}, Leonardo and {Alonso Sobrino}, Roi and {Altavilla}, Giuseppe and {Althaus}, Christian and {Alvarez Trujillo}, Luis Alonso and {Amarsi}, Anish and {Ammler-von Eiff}, Matthias and {Am{\^o}res}, Eduardo and {Andrade}, Laerte and {Antoniadis-Karnavas}, Alexandros and {Ant{\'o}nio}, Carlos and {Aparicio del Moral}, Beatriz and {Appolloni}, Matteo and {Arena}, Claudio and {Armstrong}, David and {Aroca Aliaga}, Jose and {Asplund}, Martin and {Audenaert}, Jeroen and {Auricchio}, Natalia and {Avelino}, Pedro and {Baeke}, Ann and {Bailli{\'e}}, Kevin and {Balado}, Ana and {Ballber Balaguer{\'o}}, Pau and {Balestra}, Andrea and {Ball}, Warrick and {Ballans}, Herve and {Ballot}, Jerome and {Barban}, Caroline and {Barbary}, Ga{\"e}le and {Barbieri}, Mauro and {Barcel{\'o} Forteza}, Sebasti{\`a} and {Barker}, Adrian and {Barklem}, Paul and {Barnes}, Sydney and {Barrado Navascues}, David and {Barragan}, Oscar and {Baruteau}, Cl{\'e}ment and {Basu}, Sarbani and {Baudin}, Frederic and {Baumeister}, Philipp and {Bayliss}, Daniel and {Bazot}, Michael and {Beck}, Paul G. and {Belkacem}, Kevin and {Bellinger}, Earl and {Benatti}, Serena and {Benomar}, Othman and {B{\'e}rard}, Diane and {Bergemann}, Maria and {Bergomi}, Maria and {Bernardo}, Pierre and {Biazzo}, Katia and {Bignamini}, Andrea and {Bigot}, Lionel and {Billot}, Nicolas and {Binet}, Martin and {Biondi}, David and {Biondi}, Federico and {Birch}, Aaron C. and {Bitsch}, Bertram and {Bluhm Ceballos}, Paz Victoria and {B{\'o}di}, Attila and {Bogn{\'a}r}, Zs{\'o}fia and {Boisse}, Isabelle and {Bolmont}, Emeline and {Bonanno}, Alfio and {Bonavita}, Mariangela and {Bonfanti}, Andrea and {Bonfils}, Xavier and {Bonito}, Rosaria and {Bonomo}, Aldo Stefano and {B{\"o}rner}, Anko and {Boro Saikia}, Sudeshna and {Borreguero Mart{\'\i}n}, Elisa and {Borsa}, Francesco and {Borsato}, Luca and {Bossini}, Diego and {Bouchy}, Francois and {Bou{\'e}}, Gwena{\"e}l and {Boufleur}, Rodrigo and {Boumier}, Patrick and {Bourrier}, Vincent and {Bowman}, Dominic M. and {Bozzo}, Enrico and {Bradley}, Louisa and {Bray}, John and {Bressan}, Alessandro and {Breton}, Sylvain and {Brienza}, Daniele and {Brito}, Ana and {Brogi}, Matteo and {Brown}, Beverly and {Brown}, David J.~A. and {Brun}, Allan Sacha and {Bruno}, Giovanni and {Bruns}, Michael and {Buchhave}, Lars A. and {Bugnet}, Lisa and {Buldgen}, Ga{\"e}l and {Burgess}, Patrick and {Busatta}, Andrea and {Busso}, Giorgia and {Buzasi}, Derek and {Caballero}, Jos{\'e} A. and {Cabral}, Alexandre and {Cabrero Gomez}, Juan-Francisco and {Calderone}, Flavia and {Cameron}, Robert and {Cameron}, Andrew and {Campante}, Tiago and {Campos Gestal}, N{\'e}stor and {Canto Martins}, Bruno Leonardo and {Cara}, Christophe and {Carone}, Ludmila and {Carrasco}, Josep Manel and {Casagrande}, Luca and {Casewell}, Sarah L. and {Cassisi}, Santi and {Castellani}, Marco and {Castro}, Matthieu and {Catala}, Claude and {Catal{\'a}n Fern{\'a}ndez}, Irene and {Catelan}, M{\'a}rcio and {Cegla}, Heather and {Cerruti}, Chiara and {Cessa}, Virginie and {Chadid}, Merieme and {Chaplin}, William and {Charpinet}, Stephane and {Chiappini}, Cristina and {Chiarucci}, Simone and {Chiavassa}, Andrea and {Chinellato}, Simonetta and {Chirulli}, Giovanni and {Christensen-Dalsgaard}, J{\o}rgen and {Church}, Ross and {Claret}, Antonio and {Clarke}, Cathie and {Claudi}, Riccardo and {Clermont}, Lionel and {Coelho}, Hugo and {Coelho}, Joao and {Cogato}, Fabrizio and {Colom{\'e}}, Josep and {Condamin}, Mathieu and {Conde Garc{\'\i}a}, Fernando and {Conseil}, Simon},
        title = "{The PLATO mission}",
      journal = {Experimental Astronomy},
         year = 2025,
        month = jun,
       volume = {59},
       number = {3},
          eid = {26},
        pages = {26},
          doi = {10.1007/s10686-025-09985-9},
archivePrefix = {arXiv},
       eprint = {2406.05447},
 primaryClass = {astro-ph.IM},
       adsurl = {https://ui.adsabs.harvard.edu/abs/2025ExA....59...26R}
}

@ARTICLE{Dewberry2025,
       author = {{Dewberry}, Janosz W. and {Wu}, Yanqin},
        title = "{Testing Tidal Theory Using Gaia Binaries: The Red Giant Branch}",
      journal = {\apj},
         year = 2025,
        month = may,
       volume = {984},
       number = {2},
          eid = {137},
        pages = {137},
          doi = {10.3847/1538-4357/adc37e},
archivePrefix = {arXiv},
       eprint = {2501.13929},
 primaryClass = {astro-ph.SR},
       adsurl = {https://ui.adsabs.harvard.edu/abs/2025ApJ...984..137D}
}

@ARTICLE{Beck2018,
       author = {{Beck}, P.~G. and {Mathis}, S. and {Gallet}, F. and {Charbonnel}, C. and {Benbakoura}, M. and {Garc{\'\i}a}, R.~A. and {do Nascimento}, J. -D.},
        title = "{Testing tidal theory for evolved stars by using red giant binaries observed by Kepler}",
      journal = {\mnras},
         year = 2018,
        month = sep,
       volume = {479},
       number = {1},
        pages = {L123-L128},
          doi = {10.1093/mnrasl/sly114},
archivePrefix = {arXiv},
       eprint = {1806.07208},
 primaryClass = {astro-ph.SR},
       adsurl = {https://ui.adsabs.harvard.edu/abs/2018MNRAS.479L.123B}
}

@ARTICLE{Beck2022,
       author = {{Beck}, P.~G. and {Mathur}, S. and {Hambleton}, K. and {Garc{\'\i}a}, R.~A. and {Steinwender}, L. and {Eisner}, N.~L. and {do Nascimento}, J. -D. and {Gaulme}, P. and {Mathis}, S.},
        title = "{99 oscillating red-giant stars in binary systems with NASA TESS and NASA Kepler identified from the SB9-Catalogue}",
      journal = {\aap},
         year = 2022,
        month = nov,
       volume = {667},
          eid = {A31},
        pages = {A31},
          doi = {10.1051/0004-6361/202143005},
archivePrefix = {arXiv},
       eprint = {2202.02373},
 primaryClass = {astro-ph.SR},
       adsurl = {https://ui.adsabs.harvard.edu/abs/2022A&A...667A..31B}
}

@ARTICLE{Beck2024,
       author = {{Beck}, P.~G. and {Grossmann}, D.~H. and {Steinwender}, L. and {Schimak}, L.~S. and {Muntean}, N. and {Vrard}, M. and {Patton}, R.~A. and {Merc}, J. and {Mathur}, S. and {Garcia}, R.~A. and {Pinsonneault}, M.~H. and {Rowan}, D.~M. and {Gaulme}, P. and {Allende Prieto}, C. and {Arellano-C{\'o}rdova}, K.~Z. and {Cao}, L. and {Corsaro}, E. and {Creevey}, O. and {Hambleton}, K.~M. and {Hanslmeier}, A. and {Holl}, B. and {Johnson}, J. and {Mathis}, S. and {Godoy-Rivera}, D. and {S{\'\i}mon-D{\'\i}az}, S. and {Zinn}, J.~C.},
        title = "{Constraining stellar and orbital co-evolution through ensemble seismology of solar-like oscillators in binary systems. A census of oscillating red giants and dwarf stars in Gaia DR3 binaries}",
      journal = {\aap},
         year = 2024,
        month = feb,
       volume = {682},
          eid = {A7},
        pages = {A7},
          doi = {10.1051/0004-6361/202346810},
archivePrefix = {arXiv},
       eprint = {2307.10812},
 primaryClass = {astro-ph.SR},
       adsurl = {https://ui.adsabs.harvard.edu/abs/2024A&A...682A...7B}
}

\appendix
\section{Orbital evolution equations for eccentricity}\label{app:eccentricity}
    The equations for the evolution of the eccentricity computed in \cite{EsseldeursDecin2026} are given by
    \begin{equation}
        \frac{\dd e^2}{\dd t} = \left.\frac{\dd e^2}{\dd t}\right|_{\rm wind} + \left.\frac{\dd e^2}{\dd t}\right|_{\rm tides}
    \end{equation}
    where both the contributions of the stellar wind and the tides are given by
    \begin{equation}
        \left.\left\langle\frac{\dd e^2}{\dd t}\right\rangle\right|_{\rm wind} = - 4 e^2 \frac{\dot m_1}{m_1}\langle\beta\rangle\left( \frac{1}{q} - \eta\frac{1+q}{q} - \frac{1}{2}\frac{1}{1+q}\right)
    \end{equation}
    where $\langle\rangle$ denotes orbit-averaged quantities, $\langle\beta\rangle = \beta_\mathrm{c} \frac{1}{\sqrt{1-e^2}}$, and
    \begin{equation}
        \begin{aligned}\left.\frac{\dd e^2}{\dd t}\right|_\mathrm{tides} = -\frac{5}{4\pi} \frac{m_2}{m_1} {\left(\frac{R_\star}{a}\right)}^5 \sqrt{1-e^2}\Omega_o &\sum_{m, n}\left(n\sqrt{1-e^2} - m\right)\\[-5pt]&\hspace{6pt}\times|A_{2, m, n}|^2 \Im (k^{2, m}_n)\end{aligned}\label{Eq:dedt}
    \end{equation}

\section{Kippenhahn diagram}\label{app:Kippenhahn}
\begin{figure}
        \centering
        \includegraphics[width=0.98\linewidth]{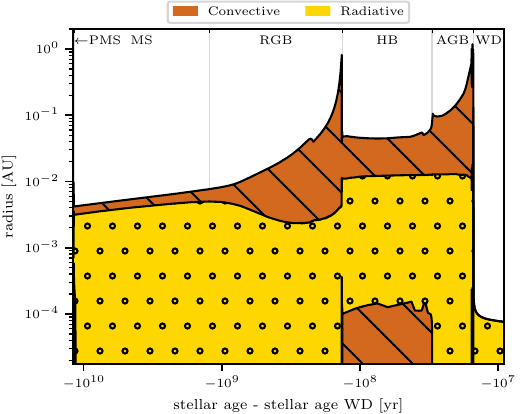}
        \vspace{-6pt}\caption{Kippenhahn diagram of the evolution of the Sun. The brown hatched regions indicate convective zones, while the yellow dotted regions indicate radiative zones. The stellar age ($x$-axis) is visualised as the time until the end of the simulation in logarithmic scale to better distinguish the different evolutionary phases.}\label{Fig:Kippenhahn}
    \end{figure}

\FloatBarrier
    For the $\eta_\text{Bl{\"o}cker}=0.05$ model the Kippenhahn diagram is shown in Fig. \ref{Fig:Kippenhahn}, indicating the different stellar evolutionary phases, as well as the convective and radiative regions. The stellar age is visualised as the time until the end of the simulation, that is, the time until the WD is cool enough to produce a luminosity of $L = 10^{-1}$ L$_\odot$. During the main sequence, the Sun has a convective envelope and a radiative core. During the RGB phase, the convective envelope deepens and the star expands. During the helium flash, the star contracts and the convective envelope recedes, creating a radiative shell around a convective core (a three-layer structure). During the AGB phase, the star expands again and the convective envelope deepens again. Finally when the star sheds its envelope, it contracts again and becomes a white dwarf.

\section{Previous tidal dissipation prescriptions}\label{app:tidaldissipationequations}
    Following \cite{Zahn1966a,Zahn1977,Hut1981,Hurley2002,Mustill2012}, and rewriting the equations in terms of the tidal Love numbers, the changes through tides can be written as
    \begin{equation}
        \Im (k^{2, 2}_2)_\mathrm{eq} = \frac{1}{27}\frac{1}{t_\text{conv}}\frac{m_\text{env}}{Gm_1^2}R_\star^3 \omega_t F(\omega_t)\ ,
    \end{equation}
    where $t_\text{conv}$ is the convective turnover timescale, $m_\text{env}$ is the mass of the convective envelope, $\omega_t$ is the tidal frequency, and $F(\omega_t)$ is a frequency-dependent factor that accounts for the interaction between turbulent convection and tidal flows \citep[see e.g.][]{Duguid2020}. 
    This relation assumes that the tidal dissipation is dominated by the equilibrium tide, as well as parametrisations for stellar properties. For the frequency-dependent factor, we use the prescription of \cite{Mustill2012}, where
    \begin{equation}
        t_\text{conv} = \left(\frac{m_\text{env} R_\text{env}^2}{\eta_F L}\right)^{1/3},\ \
        F(\omega_t) = f^\prime\min\left(1, \left|\frac{2\pi}{\omega_t c_F t_\text{conv}}\right|^{\gamma_F}\right)\ ,
    \end{equation}
    where $\eta_F$, $f^\prime$, $c_F$, and $\gamma_F$ are parameters of order unity. We adopt their parameters of $\eta_F=3$, $f^\prime=9/2$, $c_F=1$, and $\gamma_F=2$.

    A first difference with the prescriptions of \cite{Esseldeurs2024} is that in \cite{Esseldeurs2024} the dynamical tide associated with progressive internal gravity waves is also included, while in the prescriptions of \cite{Mustill2012} only the equilibrium tide is included. However, this dynamical tide is expected to be subdominant in the giant phases \citep[see][]{Beck2018,Beck2022,Beck2024,Esseldeurs2024,Dewberry2025,EsseldeursDecin2026}. A second difference comes from the different parametrisations of the equilibrium tide, where the prescriptions of \cite{Zahn1966a} evaluate the convective turnover timescale as one value for the entire convective envelope, while the prescriptions of \cite{Esseldeurs2024} evaluate the convective turnover timescale locally at each radius in the convective envelope, which leads to a more accurate evaluation of the tidal dissipation. The bulk of the convective envelope however has a relatively uniform convective turnover timescale, so the difference is not expected to be large. Lastly, the prescriptions of \cite{Esseldeurs2024} are calibrated on numerical simulations of convection of \cite{Duguid2020}\footnote{Although the simulations of \cite{Duguid2020} are currently the best in the state of the art, they still don't perfectly represent true stellar conditions. They only compute values of the Rayleigh number up to $10^3$, where although their results are relatively robust against the Rayleigh number in their regime, deviations to the prescription when adopting more realistic Rayleigh numbers (and Prandtl numbers; for the current Sun $\text{Ra} \in [10^{21}, 10^{24}]$ and $\text{Pr} \in [10^{-7}, 10^{-3}]$ \citealp{Hanasoge2016}) would not be unexpected.}, while previous studies like \cite{Mustill2012} use the \cite{Goldreich1977} $\gamma_F$ = 2 fast-tide scaling. The Earth-Sun system hovers around the transition between the fast and slow tide regime ($\omega_t t_\text{conv} \approx 1$), exactly where the difference between the two prescriptions is the largest and can change up to an order of magnitude. This makes the choice of prescription for the interaction between convective turbulence and tidal flows crucial for the orbital evolution of the Earth.

\section{Mean motion resonances}\label{app:meanmotionresonances}
\begin{figure}
        \centering
        \includegraphics[width=0.98\linewidth]{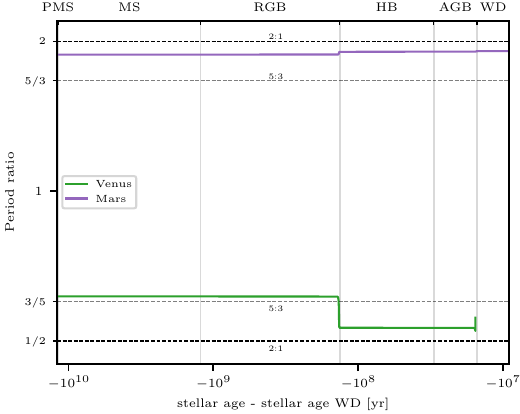}
        \vspace{-6pt}\caption{Period ratio of the Earth with respect to the closest planets, where green represents Venus and purple represents Mars.}\label{Fig:MeanMotionResonances}
    \end{figure}

    In our modeling of the orbital evolution of the inner Solar System, we do not take into account the effect of planet-planet interactions on the orbital evolution. However, as the star evolves and the planets move outward, they can cross mean motion resonances with other planets, which can lead to changes in the eccentricity and inclination of the planets, and thus affect their orbital evolution. To check if this is the case for Earth, we show in Fig. \ref{Fig:MeanMotionResonances} the period ratio of the Earth with respect to Venus and Mars during the evolved phases of the Sun. It can be seen that Earth does not cross any first-order mean motion resonance with Venus or Mars during the evolved phases of the Sun, however Venus-Earth does cross the 5:3 second-order mean motion resonance during the RGB phase. This means that the effect of planet-planet interactions on the orbital evolution of the Earth could be important during the RGB phase, but is likely negligible during the AGB phase.

\section[The mass-loss rate of L2 Pup]{The mass-loss rate of L$_2$ Pup}\label{app:mass-loss}
    The AGB star L$_2$ Pup is an important observational proxy for the Sun during the AGB phase, as its predicted initial mass is close to that of the Sun (0.98 M$_\odot$; \citealp{Kervella2016}). There are two different estimates of the mass-loss rate of L$_2$ Pup: one based on the dust emission, which gives a dust mass-loss rate of $\dot m_{1,d} = 2.5 \times 10^{-9}$ M$_\odot$ yr$^{-1}$ and a gas-to-dust ratio of 400, corresponds to a total mass-loss rate of $\dot m_1 = 1 \times 10^{-6}$ M$_\odot$ yr$^{-1}$ \citep{Haworth2018}; and one based on the CO emission, which gives a much lower mass-loss rate of $\dot m_1 = 1.2 \times 10^{-8}$ M$_\odot$ yr$^{-1}$ \citep{Hoai2022}. Both estimates take into account the effect of the disk around L$_2$ Pup and derive mass-loss rates similar to previous estimates that did not account for the disk \citep[see, e.g.,][]{Olofsson2002,Bedding2002,Jura2002,Danilovich2015}. The estimate based on the dust emission likely best traces the motion of material close to the star (i.e., within the disk), while the estimate based on the CO emission likely best traces the motion of material in the outer regions of the circumstellar environment. Both methods have their uncertainties; however, it is less clear whether the dust mass-loss rate can be directly converted into a gas mass-loss rate using a fixed gas-to-dust ratio, although this is unlikely to change the inferred value by more than an order of magnitude. Using a radius of 123 R$_\odot$ for L$_2$ Pup \citep{Kervella2014}, a mass of 0.66 M$_\odot$ \citep{Kervella2016}, and a luminosity of 1480 L$_\odot$ \citep{Uttenthaler2024}, the dust-based mass-loss rate corresponds to $\eta_\text{Bl{\"o}cker} = 2.5$, while the CO-based mass-loss rate corresponds to $\eta_\text{Bl{\"o}cker} = 0.03$. Given the large difference between the two estimates, it remains unclear which one best represents the true AGB mass-loss rate of the Sun. For this reason, we adopt $\eta_\text{Bl{\"o}cker} = 0.05$ as a reference value, as it lies within the uncertainty range of the CO-based estimate towards the dust-based estimate.

\end{document}